\documentclass[aps,pra,twocolumn]{revtex4}
\usepackage{graphicx}
\usepackage{amssymb}
\usepackage{amsmath}

\newcommand{\me}{\mathrm{e}}

\begin{document}

\title{Constructing Functional Braids for Low-Leakage Topological
 Quantum
 Computing}

\author{Haitan Xu and Xin Wan}
\email{xinwan@zimp.zju.edu.cn} \affiliation{Zhejiang Institute of
  Modern Physics, Zhejiang University, Hangzhou, 310027, China}

\begin{abstract}
  We discuss how to significantly reduce leakage errors in topological
  quantum computation by introducing an irrelevant error in phase,
  using the construction of a CNOT gate in the Fibonacci anyon model
  as a concrete example. To be specific, we construct a functional
 braid in a
  six-anyon Hilbert space that exchanges two neighboring
  anyons while conserving the encoded quantum information. The leakage
  error is $\sim$$10^{-10}$ for a braid of $\sim$100 interchanges of
  anyons. Applying the braid greatly reduces the leakage
  error in the construction of generic controlled-rotation gates.
\end{abstract}

\maketitle
\section{Introduction}
Topological quantum computing is a novel concept aiming at
preventing quantum decoherence in quantum computation on a hardware
level~\cite{kitaev03,freedman02,Preskill,dassarma07}.  The prototype
of a topological quantum computer is a two-dimensional system of
topological order with a collection of non-Abelian anyons, which can
be created, braided, fused, and measured. Quantum information is
stored globally and thus affected only by global operations like
braidings of anyons, but not by local perturbations like noises. The
leading candidates for topological states of matter are fractional
quantum Hall liquids, in which some of the fractionally charged
quasiparticles are believed to obey non-Abelian
statistics~\cite{dassarma07}. As a consequence, the architecture and
algorithms for topological quantum computers are significantly
different from conventional quantum computers such as ion-trap
quantum computers. Topological quantum computers are based on an
anyon model, shown to be equivalent to a quantum circuit
model~\cite{freedman02}. Due to its intrinsic connection to the
unitary representation of the braid groups, a perfect problem for it
is naturally the evaluation of the Jones
polynomial~\cite{freedman03,aharonov05}, which has been shown to be
intimately connected to topological quantum field theories
(TQFTs)~\cite{witten89}. TQFTs are low-energy effective theories for
topological phases of matter, in which non-Abelian anyons might
exist. The worldlines of anyons braiding in (2+1)-dimensional
space-time (with proper closure) can be regarded as knots or links.
To achieve universal topological quantum computing, one may wish to
construct a universal set of quantum gates, which are certain braids
in (2+1)-dimensional space-time. However, to solve an apparently
easier problem such as the realization of a CNOT gate is highly
nontrivial for a concrete anyon model~\cite{bonesteel05,hormozi07}.

While the discussions here may apply generically to other models
that support universal topological quantum computation, we continue
by focusing specifically on the simple Fibonacci model (equivalent
to the $SU(2)_3$ Chern-Simons-Witten theory for our purpose). Using
the simplest model, Bonesteel {\it et~al.}~\cite{bonesteel05}
explicitly showed a CNOT gate can be approximated (to a distance
$\sim$$10^{-3}$~\cite{distance}) by the composite braid of an
injection weave, an $iX$ weave, and the inverse injection weave (up
to a single-qubit rotation). The divide-and-conquer approach avoids
the difficulties of the direct search in the space of braids with at
least 6 anyons. Later, Hormozi {\it et al.}~\cite{hormozi07} also
proposed that a CNOT gate can be alternatively approximated (also to
a distance $\sim$$10^{-3}$) by the composite braid of an $F$ weave,
a $\pi$-phase weave, and the inverse $F$ weave (up to two
single-qubit rotations). These three-anyon weaves (a subset of
braids in which only one of the three anyons moves) are constructed
by brute-force search~\cite{bonesteel05} and the accuracy of the
CNOT gate can be systematically improved by the Solovay-Kitaev
algorithm~\cite{bonesteel05,hormozi07} at the expense of increasing
the braid length significantly.  The injection weave and the $F$
weave, which approximate the identity matrix and the $F$ matrix,
respectively, are examples of {\it functional braids} that might
play important roles in future topological quantum computer
architecture.

In the map of a topological quantum computer to a quantum circuit
model, qubits are encoded by several non-Abelian anyons, which, in
general, have a Hilbert space dimension larger than 2.  This means
the anyons can evolve to a noncomputational state after a braiding
involving more than one qubits. The unwanted transitions are thus
leakage errors. In the earlier works~\cite{bonesteel05,hormozi07},
the leakage error of a generic controlled-rotation gate is
determined by the error of the injection weave or the $F$ weave
($\sim$$10^{-3}$), the same as that of a generic single-qubit gate.
It is thus of great interest to reduce leakage errors for these
gates commonly used in quantum computing.

In this letter, we discuss a promising low-leakage construction of
generic controlled-rotation gates in the Fibonacci anyon model,
which is based on an {\it arbitrary} single-qubit phase gate. {\it
  Regardless of the phase}, we can use the underlying braid to
construct an {\it exchange braid} that swaps two non-Abelian anyons
in a six-anyon system without changing the quantum information
stored in the system. The freedom in the phase guarantees us to
achieve leakage errors smaller than that of a generic single qubit
of the same braid length, e.g., a leakage error $\sim$$10^{-10}$ (in
terms of distance) with no more than 100 interchanges of neighboring
anyons.  This allows us to construct two-qubit gates with extremely
low leakage errors. Other applications of the braid and the
generalization of this approach will be discussed.

\section{Fibonacci anyons}
The Fibonacci anyons emerge from the Yang-Lee model in conformal
field theories and are also speculated to exist in the $\nu = 12/5$
fractional quantum Hall liquid~\cite{read99,xia04}, as well as in
the non-Abelian spin-singlet state~\cite{ardonne99}. The Fibonacci
anyon model contains two types of anyons with topological charges
$0$ (vacuum) and $1$ (Fibonacci anyon) satisfying a nontrivial
fusion rule $1\times 1 = 0 + 1$. We denote $|(ab)_c\rangle$ for the
state in which anyons $a$ and $b$ fuse into $c$.  As in a generic
anyon model, the braiding rules for two Fibonacci anyons can be
described by the $R$-matrix and the associativity of fusion
(satisfying a pentagon relation) can be described by the
$F$-matrix~\cite{Preskill}.

\begin{figure}
 \begin{center}
 \includegraphics[width=7cm]{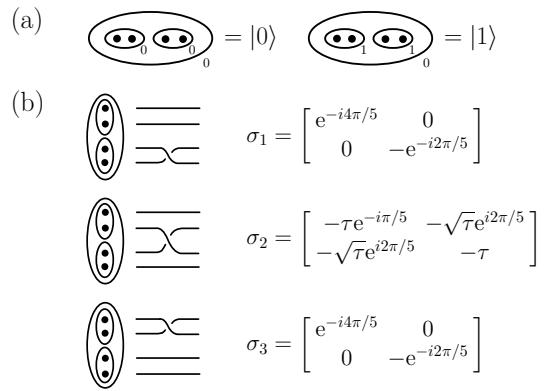}
 \end{center}
 \caption{\label{fig:qubit} (a) Basis states for the two-dimensional
   Hilbert space (a qubit) spanned by two pairs of anyons with total
   topological charge 0, where the black dots denote the Fibonacci
   anyons in the Fibonacci model. (b) Braiding matrices for the qubit,
   where $\tau=(\sqrt{5}-1)/2$.  The anyons are labeled 1 to 4 from
   left to right in (a) and from bottom to top in (b).}
 \end{figure}

We use two pairs of Fibonacci anyons with total charge 0 to encode
one qubit of information. As illustrated in
Fig.~\ref{fig:qubit}~(a), the two pairs must have the same total
charge 0 or 1, spanning a two-dimensional Hilbert space, or a qubit.
We can write down the basis as $|0\rangle=|((11)_0(11)_0)_0\rangle$
and $|1\rangle=|((11)_1(11)_1)_0\rangle$.  We note that,
alternatively, we can use a set of three anyons with total charge 1,
such that $|0\rangle=|((11)_01)_1\rangle$ and
$|1\rangle=|((11)_11)_1\rangle$. The two encoding schemes are
equivalent since $|((11)_a(11)_a)_0\rangle =
|(((11)_a1)_11)_0\rangle$, which is just $|((11)_a1)_1\rangle$ in
the three-anyon encoding scheme, as explained, e.g., in
Ref.~\cite{hormozi07}. The braiding of neighboring anyons in the
qubit can be represented by $2 \times 2$ matrices, as shown in
Fig.~\ref{fig:qubit}~(b), obtained from $F$- and $R$-matrices. There
are only two independent matrices, since $\sigma_1=\sigma_3$, so, in
practice, we can leave, {\it e.g.}, the bottom anyon intact.
Therefore, we can use 4 elementary matrices $\sigma_{2}^{\pm1}$ and
$\sigma_{3}^{\pm1}$ (equivalent to the braiding matrices in the
three-anyon encoding scheme) to construct any single-qubit gate,
since the Fibonacci anyon model supports universal topological
quantum computation.

\section{ Single-qubit gates} A generic single-qubit gate can be
represented by a $2\times2$ unitary matrix, which can be written in
the form
\begin{equation}
\label{eq:matrix} \me^{i\alpha}\left[
 \begin {array}{cc}
 \sqrt{1-b^2}\me^{-i\beta} &  b\me^{i\gamma}\\
 -b\me^{-i\gamma} & \sqrt{1-b^2}\me^{i\beta}
\end{array}\right],
\end{equation}
where $b$, $\alpha$, $\beta$ and $\gamma$ are all real parameters.
Apart from the overall phase factor, one needs 3 parameters $b$,
$\beta$ and $\gamma$ to specify the matrix. To search for any
desired gate, we implement the brute-force
algorithm~\cite{bonesteel05} with some modifications. In particular,
we store the intermediate matrix products, as we generate longer and
longer braids, ordered by the parameters $b$, $\beta$, and $\gamma$
in Eq.~(\ref{eq:matrix}). When we reach our searching limit (on a
desktop PC), we look up in the saved list an optimal braid that can
be combined to approach the target gate. In computer algorithms,
this is equivalent to a bidirectional search~\cite{thanknick}. We
find, in general, single-qubit gates [such as the $iX$ gate in
Fig.~\ref{cphase}(b)] can be approximated by braids of about 100
interchanges of neighboring anyons to a distance $\sim$$10^{-6}$,
sufficiently small for the application of quantum error-correcting
code.

\begin{figure*}
  \begin{center}
    \includegraphics[width=17cm]{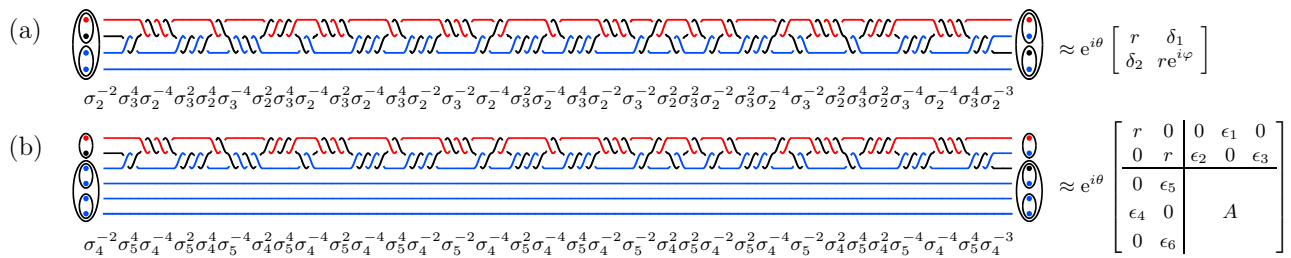}
  \end{center}
  \caption{\label{fig:diagonalgate} (Color online)  (a) A braid with 99
 anyon
    interchanges that approximates a phase gate, where
    $|\delta_1|=|\delta_2|=\sqrt{1-r^2}\simeq6.2 \times 10^{-10}$. (b)
    The exchange braid in the six-anyon braid space, in which
    $\sigma_4$ and $\sigma_5$ are the corresponding $5 \times 5$
    braiding matrices. Note that the sequence of interchanges for the
    top 3 anyons is exactly the same as in (a). The phase factor
    $\me^{i\theta} $ inherited from (a) is not important and can be
    canceled out by the inverse exchange braid. The upper $2 \times 2$
    block is exactly proportional to the identity matrix, and
    $1-r\approx 2\times10^{-19}$. The lower $3 \times 3$ block $A$ (in
    the noncomputational space) is irrelevant, if the quantum state
    remains in the computational space. The leakage to the
    noncomputational space spanned by $\vert (((11)_1 (11)_0)_1
    (11)_1)_0 \rangle$, $\vert (((11)_0 (11)_1)_1 (11)_1)_0 \rangle$
    and $\vert (((11)_1 (11)_1)_1 (11)_1)_0 \rangle$ is determined by
    $|\epsilon_i| \leqslant 6.2 \times 10^{-10}$ for $i=1$-$6$.}
\end{figure*}

One may notice that when $b = 0$, $\gamma$ drops out automatically.
The resulting diagonal matrix is a phase gate in the quantum
computing language.  Compared with the identity matrix targeted by
the injection weave~\cite{bonesteel05}, an additional error in phase
is introduced. However, if the phase error is irrelevant (as we will
see), the leakage error due to non-zero off-diagonal matrix elements
can be greatly reduced. This is because, with only one parameter $b$
to target, some diagonal matrices can be approximated with much
higher accuracy within a certain braid length.  As an example, we
show in Fig.~\ref{fig:diagonalgate}(a) a sequence of 99 elementary
braids (or a braid of length 99) that approaches a phase gate with
off-diagonal matrix elements of order $10^{-10}$ ({\it i.e. more
than 3 orders of
  magnitude improvement} over the generic case). Note the braid
exchanges the positions of the two center anyons, while rotating the
qubit only by a phase. In the following, we show this plain-looking
braid can facilitate topological quantum computing with remarkably
low leakage error.

\section{From one to two qubits} Let us create an auxiliary pair of
Fibonacci anyons with total charge 0 next to the two pairs that
encode a qubit. The whole Hilbert space of the 6 Fibonacci anyons
with total charge 0 is 5-dimensional. The additional 3 basis states
have a total charge 1 for the 4 encoding anyons, thus are
noncomputational basis states. In the enlarged space, we note the
computational basis states for the 6 anyons can be rewritten as
$\vert (((11)_a (11)_a)_0 (11)_0)_0 \rangle = \vert (((11)_a 1)_1 (1
(11)_0)_1)_0 \rangle$. Since a set of 3 nontrivial anyons with total
charge 1 is equivalent to a set of 4 with total charge 0 in terms of
computing basis and corresponding braiding matrices, the six-anyon
system can thus be regarded as two qubits (with total charge 0),
except that the second (right) qubit is in the definite state of
$\vert(1 (11)_0)_1 \rangle$. Suppose we now implement the braid in
Fig.~\ref{fig:diagonalgate}(a) to the second qubit, as shown in
Fig.~\ref{fig:diagonalgate}(b). We obtain just a phase factor since
the braid is represented by a diagonal matrix in the single-qubit
representation.  As a result, the positions of the fourth and fifth
anyons are exchanged, but the quantum information stored in the
qubit remains intact.

In the full 5-dimensional space, the braiding of the second set of
anyons leads to a $2 \times 2$ (computational) and a $3 \times 3$
(noncomputational) matrices. They are decoupled (block-diagonal) to
the order of $10^{-10}$, the same order as the off-diagonal matrix
elements of the single-qubit phase gate in
Fig.~\ref{fig:diagonalgate}(a). The $2 \times 2$ submatrix is {\it
  exactly} proportional to the identity matrix. The phase we
introduced as an error drops out and is indeed irrelevant.  The
braid in the 6-anyon system shown in Fig.~\ref{fig:diagonalgate}(b)
is thus dubbed {\it the exchange braid}, since its sole function is
to exchange one anyon in the encoded qubit with another in the
auxiliary pair without distorting the quantum information encoded in
the qubit (though with a leakage error $\sim$$10^{-10}$).  The
exchange braid has many applications. For example, it can facilitate
topological quantum computing with only a few mobile anyons.
Compared to the work published earlier~\cite{simon06} based on the
injection weave (with a distance $\sim$$10^{-3}$ and a length of
48)~\cite{bonesteel05}, with the exchange braid approach, we can
realize the function of injection (with a phase to be canceled by
its inverse) with much lower leakage errors. Since we introduce
pairs of anyons with total topological charge 0, it is also possible
to insert the pairs at any location.

\section{Controlled gates} 
\begin{figure*}
 \begin{center}
 \includegraphics[width=17cm]{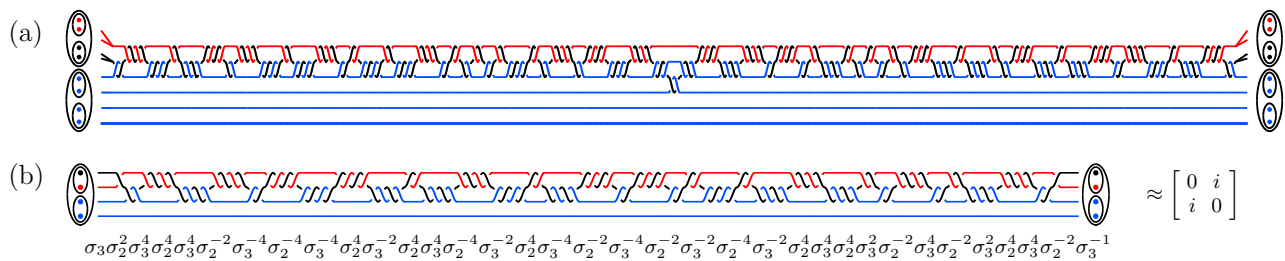}
 \end{center}
 \caption{\label{cphase} (Color online) (a) A braid that approximates a
   controlled-$\sigma_3^2$ gate to a distance
   $\simeq1.2\times10^{-9}$.  The control (top) qubit can be
   treated as two composite anyons and each composite anyons can move
   as a single anyon. In the six-anyon braid space, we apply the
   exchange braid [see Fig.~\ref{fig:diagonalgate}(b)], $\sigma_3^2$
   (two counterclockwise interchanges of the third and the fourth
   anyons from the bottom), and the inverse exchange braid. If the
   control qubit is in the state $|0\rangle$, the encoded qubit
   remains unchanged; otherwise, a $\sigma_3^2$ rotation is performed.
   (b) A braid with 108 interchanges that approximates an $iX$ gate to
   a distance $\simeq1.5\times10^{-6}$. Replacing the $\sigma_3^2$
   sandwiched in (a) by the $iX$ braid, we obtain a braid that
 approximates a
   CNOT gate (up to a single-qubit rotation) to a distance $\sim
   10^{-6}$.}
\end{figure*}

We can imagine the two Fibonacci anyons of
the auxiliary pair are themselves two pairs of Fibonacci anyons,
with total charge 1 each. The four anyons thus form a second qubit
in the state $\vert ((11)_1 (11)_1)_0 \rangle$. Let us treat the two
subpairs as single anyons and perform the exchange braid to the
two-qubit system.  The result is one composite anyon exchanges its
position with the neighboring anyon in the encoded qubit. Suppose we
then perform two consecutive $\sigma_3$ braidings, followed by the
inverse of the exchange braid in the end. The composite braid
performs a $\sigma_3^2$ to the encoded qubit. On the other hand, we
can also encode the second qubit in the state $\vert ((11)_0
(11)_0)_0 \rangle$. Since the two composite anyons in the auxiliary
pair each have trivial total charge, the composite braid thus makes
no change to the encoded qubit. Combining the two scenarios, we
conclude the composite braid (of length 200) for the 8-anyon system,
as shown in Fig.~\ref{cphase}(a), approximates the
controlled-$\sigma_3^2$ operation to a distance $\sim$$10^{-9}$.

In the same fashion, we can construct any controlled-rotation gate
by replacing the $\sigma_3^2$ braid by one that approximates the
appropriate single-qubit gate.  For example, we find a braid of
length 108 that approximates the $iX$ gate to a distance $\simeq 1.5
\times 10^{-6}$ as shown in Fig.~\ref{cphase}(b).  Sandwiching the
$iX$ braid between the exchange braid and its inverse, we obtain a
braid (of length 306) that approximates a CNOT gate (up to a
single-qubit rotation) to a distance $\simeq 1.5\times 10^{-6}$ with
a leakage error $\sim$$10^{-9}$. One can also achieve a more
accurate CNOT gate by {\it only} improving the $iX$ gate (to
distance $\sim$$10^{-9}$) using, {\it e.g.}, the Solavay-Kitaev
algorithm~\cite{kitaev99}.

\section{ Discussion}  
\begin{figure}[htb]
 \begin{center}
 \includegraphics[width=8cm]{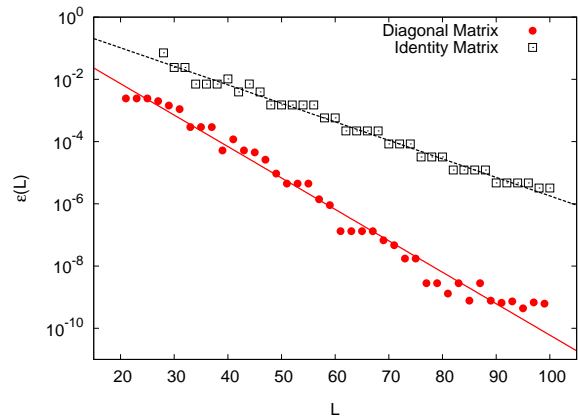}
 \caption{\label{length_vs_error} (Color online) Distance versus braid
   length in approaching the identity matrix (dots) or any phase gate,
   {\it i.e.}, unitary diagonal matrix (squares).  The decays can be
   fit by $1.6 e^{-L/7.3}$ (dashed line) and $0.76 e^{-L/4.3}$ (solid
   line), respectively.  The significant reduction in distance due to
   the freedom in phase is key to the low-leakage realization of
   topological quantum computation based on the exchange braid.}
 \end{center}
\end{figure}

To summarize, we introduce a scheme to construct
 a
functional braid that can be interpreted either as a phase gate
(with off-diagonal matrix elements $\sim$$10^{-10}$) or as an
exchange braid, which swaps two neighboring anyons without affecting
the encoded quantum information (except for a leakage error
$\sim$$10^{-10}$).  The intriguing equivalence between the single-
and two-qubit constructions holds the key to the remarkable
reduction in leakage errors.  We illustrate this in
Fig.~\ref{length_vs_error}, where we plot the distance we can
approach in searching for the identity matrix, as well as the
distance for phase gates (with arbitrary phases). We point out the
curve for the identity matrix (dashed line) extends the earlier
approach (up to $L = 48$ as in Fig.~7 of
Ref.~\onlinecite{hormozi07}) with the algorithm improvement, which
reduces the leakage errors by 3 orders of magnitude by doubling the
braid length (achieved with similar computing power). Remarkably,
the phase gate based approach further reduces the leakage errors by
another 3 orders of magnitude {\it without extra cost on the braid
  length}.  Gain in the approach happens at all braid lengths and is
based mathematically on the increase of the target space from 0 to 1
dimension, which leads to a faster exponential decay in distance as
illustrated in Fig.~\ref{length_vs_error} (quantitatively, a factor
of $\sim$1.7 reduction in decay length).

This study reveals an interesting aspect of the topological quantum
computation: {\it errors (in leakage) can be reduced by
  introducing an additional error (in phase)}. Based on
the guideline, one can engineer more functional braids for Fibonacci
anyons~\cite{xu07}, {\it e.g.}, one that approximates a generalized
NOT gate (an off-diagonal matrix) or a generalized Hadamard gate to
high accuracy. These functional braids can readily find applications
in low-leakage topological quantum algorithms.  While we have used
the simple Fibonacci anyon model as an example to illustrate our
ideas to achieve low-leakage topological quantum computing, such
functional braids can be constructed in similar fashions in a
generic anyon model that supports universal topological quantum
computing.

\section*{ACKNOWLEDGEMENTS}
We thank Nick Bonesteel for helpful discussion and a careful reading
of an earlier version of the manuscript. X.W. benefits greatly from
the lectures on quantum topology by Zhenghan Wang. This work is
supported by the NSFC Grant No.~10504028, the PCSIRT (Project
No.~IRT0754), and, partially, by SRF for ROCS, SEM.


\begin{thebibliography}{99}
\bibitem{distance} As in Ref.~\onlinecite{bonesteel05}, the distance
  between two matrices (or quantum gates) $A$ and $B$ is defined to be
  the square root of the highest eigenvalue of $(A-B)^\dagger(A-B)$.
\end{thebibliography}
\end{document}